\documentclass[12pt]{article}
\usepackage{amssymb, amsmath, amsopn, amsthm}

\usepackage{bm,amsfonts,array,calc,rotating}
\usepackage{epsfig}
\usepackage{graphicx}
\usepackage{color}
\usepackage[colorlinks=false]{hyperref}

\global\def\draftcontrol{0}

   \def\versionno{Supersymmetry breaking, heterotic strings and fluxes}

\catcode`\@=11

\expandafter\ifx\csname
draftcontrol\endcsname\relax\global\def\draftcontrol{0} \fi

{\count255=\time\divide\count255 by 60
\xdef\hourmin{\number\count255} \multiply\count255
by-60\advance\count255 by\time
\xdef\hourmin{\hourmin:\ifnum\count255<10 0\fi\the\count255}}
\def\draftdate{\number\month/\number\day/\number\year\ \ \ \hourmin }

\newcommand\makepapertitle{\par

  \begingroup
    \renewcommand\thefootnote{\@fnsymbol\c@footnote}%
    \def\@makefnmark{\rlap{\@textsuperscript{\normalfont\@thefnmark}}}%
    \long\def\@makefntext##1{\parindent 1em\noindent
            \hb@xt@1.8em{%
                \hss\@textsuperscript{\normalfont\@thefnmark}}##1}%
     \newpage
     \global\@topnum\z@   
     \@makepapertitle
     \thispagestyle{empty}\@thanks
  \endgroup
  \setcounter{footnote}{0}%
  \global\let\thanks\relax
  \global\let\makepapertitle\relax
  \global\let\@makepapertitle\relax
  \global\let\@thanks\@empty
  \global\let\@author\@empty
  \global\let\@date\@empty
  \global\let\@title\@empty
  \global\let\title\relax
  \global\let\author\relax
  \global\let\date\relax
  \global\let\and\relax
  \def\version{\let\version\@version\@gobble}
}
\def\@makepapertitle{%
  \newpage
   \ifnum\draftcontrol=1{}
   \version\versionno
   \vskip 5em%
   \else
   \hfill\hbox to 3cm {\parbox{4cm}{\@pubnum}\hss}%
   \vskip 5em%
   \fi
   \begin{center}%
   \let \footnote \thanks
      {\hskip -0\textwidth \hbox to 1\textwidth%
        {\centerline{\Large\bf{\noindent\@title}}}}%
     \vskip 2em%
     {\normalsize
       \lineskip .5em%
       \begin{tabular}[t]{c}%
         \@author
       \end{tabular}\par}%
     \vskip 1em%
     {\@bstract}%
     \end{center}%
     \vfill
     \@date%
     \vskip 1.5em%
     \noindent
     \rule{12em}{.02em}\par\noindent
     \@email%
   \par
}

\gdef\@pubnum{}
\def\pubnum#1{%
  \gdef\@pubnum{#1}}

\gdef\@bstract{}
\def\Abstract#1{%
  \gdef\@bstract{%
   \parbox{\textwidth-0pc}{%
   \centerline{\bf Abstract}\penalty1000
   \noindent
   \renewcommand\baselinestretch{1.0}
   {#1}}}
}

\gdef\@email{}
\def\email#1{%
   \gdef\@email{%
  {\small Email: {\tt #1}}}
}

\def\ps@paper{\let\@mkboth\@gobbletwo%
     \ifnum\draftcontrol=1
        \def\@oddfoot{\hbox to \textwidth{\tiny \versionno \hfil\tiny\draftdate}%
        \hskip -\textwidth \hbox to \textwidth{\hfil\rm\thepage\hfil}}%
     \else\def\@oddfoot{\hbox to \textwidth{\hfil\rm\thepage\hfil}}
     \fi
     \let\@evenfoot\@oddfoot
}

\def\body{\clearpage
          \pagestyle{paper}
        }

\def\@version#1{\ifnum\draftcontrol=1
\typeout{}\typeout{#1}\typeout{}
\vskip3mm\centerline{\hbox{\fbox{\normalsize{\tt DRAFT -- #1 -- }
                   {\draftdate}}}}\vskip3mm
\fi}
\let\version\@version
\long\def\eqlabel#1{\ifnum\draftcontrol=1
                    \tag@false  
                    \tag*{(\theequation) \hbox to -0.2cm{\hspace{0cm}\small{#1}\hss}}
                    \refstepcounter{equation}
                    \edef\@currentlabel{\theequation}
                    \ltx@label{#1}          
                    \else
                    \label{#1}
                    \fi
                    }
\let\st@bibitem\@bibitem
\let\st@lbibitem\@lbibitem
\ifnum\draftcontrol=1
  \def\@bibitem#1{%
    \st@bibitem{#1}\a@@label{#1}\ignorespaces}
  \def\@lbibitem[#1]#2{%
    \st@lbibitem[#1]{#2}\a@@label{#2}\ignorespaces}
  \def\a@@label#1{%
    \gdef\a@lab{\smash{\normalfont\small#1}}
    \ifvmode
      \if@inlabel
        \global\setbox\@labels\hbox{%
          \llap{\a@lab\let\a@lab\relax
                \kern\@totalleftmargin\kern\marginparsep}%
          \box\@labels}%
      \fi
    \fi}
\fi

\renewcommand\baselinestretch{1.25}
\renewcommand\section{\@startsection {section}{1}{\z@}%
                                   {-3.5ex \@plus -1ex \@minus -.2ex}%
                                   {2.3ex \@plus.2ex}%
                                   {\normalfont\large\bfseries}}
\renewcommand\subsection{\@startsection{subsection}{2}{\z@}%
                                   {-3.25ex\@plus -1ex \@minus -.2ex}%
                                   {1.5ex \@plus .2ex}%
                                   {\normalfont\normalsize\bfseries}}
\renewcommand\subsubsection{\@startsection{subsubsection}{3}{\z@}%
                                   {-3.25ex\@plus -1ex \@minus -.2ex}%
                                   {1.5ex \@plus .2ex}%
                                   {\normalfont\normalsize\it}}
\renewcommand\paragraph{\@startsection{paragraph}{4}{\z@}%
                                   {-3.25ex\@plus -1ex \@minus -.2ex}%
                                   {1.5ex \@plus .2ex}%
                                   {\normalfont\normalsize\bf}}
\renewcommand\subparagraph{\@startsection{subparagraph}{5}{\z@}%
                                   {-1.25ex\@plus -1ex \@minus -.2ex}%
                                   {0ex \@plus .2ex}%
                                   {\normalfont\normalsize\it}}

\numberwithin{equation}{section}

\long\def\@makecaption#1#2{%
  \vskip\abovecaptionskip
  \sbox\@tempboxa{{\bf #1:} #2}%
  \ifdim \wd\@tempboxa >\hsize
    {\small\bf #1:} {\small #2}\par
  \else
    \global \@minipagefalse
    \hb@xt@\hsize{\hfil\box\@tempboxa\hfil}%
  \fi
  \vskip\belowcaptionskip}

\renewcommand*\l@section[2]{%
  \ifnum \c@tocdepth >\z@
    \addpenalty\@secpenalty
    \addvspace{.5em \@plus\p@}%
    \setlength\@tempdima{1.5em}%
    \begingroup
      \parindent \z@ \rightskip \@pnumwidth
      \parfillskip -\@pnumwidth
      \leavevmode \bfseries
      \advance\leftskip\@tempdima
      \hskip -\leftskip
      #1\nobreak\hfil \nobreak\hb@xt@\@pnumwidth{\hss #2}\par
    \endgroup
  \fi}
\renewcommand*\l@subsection{\addvspace{.0em \@plus\p@}\@dottedtocline{2}{1.5em}{2.3em}}
\renewcommand*\l@subsubsection{\addvspace{-.2em \@plus\p@}\@dottedtocline{3}{3.8em}{3.2em}}

\definecolor{refcol}{rgb}{0.2,0.2,0.8}
\definecolor{eqcol}{rgb}{.6,0,0}
\definecolor{purple}{cmyk}{0,1,0,0}

\gdef\@citecolor{refcol} \gdef\@linkcolor{eqcol}
\def\colorlinkspurple{\gdef\@urlcolor{purple}}
\def\colorlinksblue{\gdef\@urlcolor{blue}}
\def\colorlinksred{\gdef\@urlcolor{red}}

\def\revise#1       {\raisebox{-0em}{\rule{3pt}{1em}}%
                     \marginpar{\raisebox{.5em}{\vrule width3pt\
                     \vrule width0pt height 0pt depth0.5em
                     \hbox to 0cm{\hspace{0cm}{%
                     \parbox[t]{4em}{\raggedright\footnotesize{#1}}}\hss}}}}

\def\a{\alpha}

\def\d{\delta}
\def\ep{\epsilon}
\def\ve{\varepsilon}

\def\k{\kappa}
\def\l{\lambda}
\def\m{\mu}
\def\n{\nu}

\def\s{\sigma}

\def\ph{\phi}

\def\c{\chi}

\def\G{\Gamma}

\def\O{\Omega}
\def\o{\over}
\def\p{\partial}

\def\p{\partial}

\def\inbar{\,\vrule height1.5ex width.4pt depth0pt}
\def\IC{\relax\hbox{$\inbar\kern-.3em{\rm C}$}}
\def\IN{\relax{\rm I\kern-.18em N}}
\def\IR{\relax{\rm I\kern-.18em R}}
\font\cmss=cmss10 \font\cmsss=cmss10 at 7pt
\def\IZ{\relax\ifmmode\mathchoice
{\hbox{\cmss Z\kern-.4em Z}}{\hbox{\cmss Z\kern-.4em Z}}
{\lower.9pt\hbox{\cmsss Z\kern-.4em Z}} {\lower1.2pt\hbox{\cmsss
Z\kern-.4em Z}}\else{\cmss Z\kern-.4em Z}\fi}
\def\tria{\triangleright}

\def\half{\frac{1}{2}}
\def\a{\alpha}

\def\be{\begin{equation}}
\def\beq{\begin{eqnarray}}

\def\d{\delta}
\def\ee{\end{equation}}
\def\eeq{\end{eqnarray}}
\def\ep{\epsilon}
\def\ve{\varepsilon}

\def\k{\kappa}
\def\l{\lambda}

\def\m{\mu}
\def\n{\nu}

\def\s{\sigma}

\def\ph{\phi}

\def\c{\chi}

\def\G{\Gamma}

\def\O{\Omega}
\def\o{\over}

\def\p{\partial}

\def\p{\partial}

\def\a{\alpha}
\def\G{\Gamma }

\def\e{\varepsilon}

\def\inbar{\,\vrule height1.5ex width.4pt depth0pt}
\def\tria{\triangleright}

\def\half{\frac{1}{2}}
\def\o{\over}

\def\pa{\partial}

\def\half{\frac{1}{2}}
\def\fourth{\frac{1}{4}}

\def\ee           {{\it e}}

\def\tr           {{\rm tr}}

\def\tria{$\triangleright$}

\def\pb{\phi_{\rm B}}
\def\ph{\phi_{\rm h}}

\newcommand{\paslash}{\ensuremath \raisebox{0.025cm}{\slash}\hspace{-0.25cm}\partial\/}

\def\sqr#1#2{{\vcenter{\vbox{\hrule height.#2pt
 \hbox{\vrule width.#2pt height#1pt \kern#1pt
 \vrule width.#2pt}\hrule height.#2pt}}}}

\newcommand{\C}[1]{$(\ref{#1})$}

\catcode`\@=12

\begin{document}

\pubnum{MIFP-09-14\\ NSF-KITP-09-30}

\title{Supersymmetry breaking, heterotic strings and fluxes}

\date{April 2009}

\author{\\[.5cm]Katrin Becker, Chris Bertinato, Yu-Chieh Chung and Guangyu Guo \
\\[.2cm] \it Department of Physics, Texas A\&M University, \\ \it College Station, TX 77843, USA\\ [1.5cm]}

\Abstract{In this paper  we consider compactifications of heterotic
strings in the presence of background flux. The background metric is
a T$^2$ fibration over a K3 base times four-dimensional Minkowski
space. Depending on the choice of three-form flux different amounts
of supersymmetry are preserved (N=2,1,0). For supersymmetric
solutions unbroken space-time supersymmetry determines all
background fields except one scalar function which is related to the
dilaton. The heterotic Bianchi identity gives rise to a differential
equation for the dilaton which we discuss in detail for solutions
preserving an N=2 supersymmetry. In this case the differential
equation is of Laplace type and as a result the solvability is
guaranteed. }

 \email{kbecker, cbertinato, ycchung,
guangyu@physics.tamu.edu}

\makepapertitle

\body


\vskip 1em

\newpage

\section{Introduction}

In this paper we study different aspects of string theory
compactifications in the presence of background flux. Our main focus
is the heterotic string compactified to four dimensions with
background NS three-form ${\cal H}$. Compactifications of heterotic
strings are interesting, both from the phenomenological and from the
mathematical point of view. In general, unbroken supersymmetry
requires the background in a compactitification of the heterotic
string to be a complex space with an hermitian metric but does not
require the space to be K\"ahler. If the background is non-K\"ahler
the NS three-form is non-vanishing and the deviation from
``K\"ahlerity'' is measured by the flux. On these spaces there
exists a globally defined two-form $J$ which is no longer closed.
Its derivative is rather related to the flux by
\cite{Hull:1985jv}\cite{S:Superstrings}
\begin{equation}
{\cal H} = i (\partial- \bar \partial ) J .
\end{equation}
While the K\"ahler case, like Calabi-Yau and torus
compactifications, have been intensively studied much less is known
about the generic case in which the compactification manifold is no
longer K\"ahler. From the mathematical point of view algebraic
geometry techniques are still missing even though some progress has
been made in describing these spaces with an explicit metric
\cite{Becker:2009df}.

We will consider torsional heterotic backgrounds which are a T$^2$
fibered over a K3 base. Depending on the choice of flux different
amounts of four-dimensional supersymmetry are preserved. While
solutions preserving an N=2,1 supersymmetry have been discussed
before in the literature, starting with ref.\cite{DRS:M-theory} (see
in particular \cite{Becker:2002sx}\cite{Becker:2006et}), the
supersymmetry breaking solutions are new. We explicitly check that
the backgrounds solve the equations of motion. For solutions
preserving an N=1,0 supersymmetry we check this at the SUGRA level.
While for solutions preserving an N=2 supersymmetry we show how to
solve the equations of motion including the first $\a'$ correction.
The spinor equations determine the background except one scalar
function related to the dilaton. The Bianchi identity for ${\cal H}$
gives rise to a differential equation for this scalar function which
is of Laplace type.

We start by discussing, and mostly reviewing, flux compactifications
of type IIB string theory on K3$\times$T$^2$ orientifolds (see for
example
refs.\cite{DRS:M-theory}\cite{TT:Compactification}\cite{Moore:2004fg}).
Depending on the choice of flux the solutions preserve an N=2,1,0
supersymmetry in four dimensions. The backgrounds solve the
equations of motion and in the supersymmetric case the spinor
equations. We check this to the leading order in $\alpha'$, {\it
i.e.} in the SUGRA approximation.

This type IIB background is dual to an heterotic SUGRA background
with non-vanishing ${\cal H}$-flux and non-trivial dilaton. The
duality chain has been described explicitly in ref.
\cite{DRS:M-theory} based on earlier work by Sen
\cite{Sen:1996vd}\cite{Sen:1997gv}. The heterotic background metric
is a T$^2$ fibred over a K3 base. We are interested in analyzing
$\alpha'$ corrections to the heterotic SUGRA background. Even though
the heterotic and type IIB backgrounds are related by duality we
will not use duality to obtain the $\alpha'$ corrected heterotic
background. Rather we will follow a different route and construct
the $\alpha'$ corrected background directly on the heterotic side.
The reason being that the present knowledge about the relevant
interactions on the world-volume of D$p$-branes and O-planes is
insufficient. So for example, the anomalous couplings described in
refs.
\cite{Green:1996dd}\cite{Morales:1998ux}\cite{Scrucca:1999uz}\cite{Scrucca:2000ae}
\cite{Craps:1998fn}\cite{Stefanski:1998yx}
\cite{Craps:1998tw} are not compatible with T-duality and additional
dependence on NS-NS and R-R fields are required to obtain
world-volume actions compatible with T-duality. On the heterotic
side, on the other hand, the action and supersymmetry
transformations are known to all relevant orders. The low-energy
effective action of the heterotic string to $O(\alpha'^3)$ has been
constructed by Bergshoeff and de Roo by supersymmetrizing the
Chern-Simons term \cite{Bergshoeff:1989de}. Our goal is to construct
the background which solves the $\alpha'$ corrected equations of
motion.

The duality between type IIB and heterotic flux backgrounds can be
explicitly checked at the level of SUGRA but beyond leading order
the duality map makes predictions about higher derivative
corrections to the world-volume action describing D$p$-branes in
type IIB theories. Thus, duality, while not being fundamental to
construct heterotic flux backgrounds, can be used as a tool to learn
about D$p$-brane effective actions. Furthermore duality to type IIB
allows us to speculate about the effect of fluxes on
four-dimensional phenomenology arising from heterotic strings in
generic backgrounds. So for example, the non-trivial heterotic
dilaton profile plays a role similar to the warp factor in type IIB
compactifications and in regions with strong warping could give rise
to large hierarchies.

In section 2 we present the type IIB flux backgrounds. To set up our
notation we review the low-energy effective `action' in section 2.1
and derive the equations of motion of type IIB SUGRA in section 2.2.
In section 2.3 we present the background which solves the equations
of motion of type IIB SUGRA and check the amount of four-dimensional
supersymmetry preserved by the different backgrounds in section 2.4.
Taking the type IIB background as a starting point we proceed in
section 3 to construct the heterotic flux background. To set up the
notation we review in section 3.1 the heterotic effective action to
$O(\alpha')$ and in section 3.2 we derive the corresponding
equations of motion. In section 3.3 we present the background and
show that it solves the SUGRA equations of motion. In section 4 we
discuss the $\a'$ corrected background. We start by presenting
explicit results for $\tr (R\wedge R)$ which are needed to solve the
Bianchi identity and Einstein equation. In section 4.1 we review the
proof that $\tr (R \wedge R)$ is a four-form of type (2,2) to
leading order in $\a'$, a condition which is needed for the
solvability of the Bianchi identity. In section 4.3, focusing on
solutions with N=2 supersymmetry, we show how to construct the
background which solves the $\a'$ corrected Bianchi identity and
supersymmetry transformations.

\section{Type IIB SUGRA background}

In this section we review type IIB flux backgrounds in which the
space-time metric is a warped product of flat 4d Minkowski space and
a K3$\times$T$^2$ orientifold (see
refs.\cite{DRS:M-theory}\cite{TT:Compactification}\cite{Moore:2004fg}
\cite{Gubser:2000vg}\cite{Grana:2000jj}). To set up the notation we
start summarizing our conventions for the type IIB SUGRA `action'
together with the corresponding equations of motion. Then we
summarize the solutions preserving different amounts of
four-dimensional supersymmetry. The analysis is done at the level of
SUGRA, {\it i.e.} without taking actions describing brane sources
into account.
\subsection{The action}

The bosonic part of the type IIB supergravity `action' in the 10d
string frame is
\begin{equation} \label{8actioniib}
{S} = {S}_{\rm NS}+{S}_{\rm R}+{S}_{\rm CS}.
\end{equation}
Here $S_{\rm NS}$ is
\begin{equation} \label{SNS}
{S}_{\rm NS} = {1\o 2\k^2} \int d^{10} x \sqrt{-g}\,
e^{-2 \pb } \left[ R +4 (\partial\pb )^2
 - {1\o 2}
|H_3|^2\right],
\end{equation}
while the parts of the action describing the massless R-R sector
fields are given by
\begin{equation}
{S}_{\rm R} = -{1\o 4 \k^2}\int d^{10} x \sqrt{-g} \left( |F_1|^2
+|\widetilde F_3|^2 + {1\o 2 } |\widetilde F_5|^2\right),
\end{equation}
\begin{equation}
{S}_{\rm CS} = {1\o 4 \k^2} \int C_4 \wedge H_3 \wedge F_3.
\end{equation}
In these formulas $F_{n+1} = d C_n$, $H_3 = d B_2$ and
\begin{equation}
\widetilde F_3 = F_3 - C_0 \,H_3
\end{equation}
\begin{equation} \label{tildeF5}
\widetilde F_5 = F_5 - {1\o 2} C_2 \wedge H_3 + {1\o 2} B_2 \wedge
F_3.
\end{equation}

\subsection{Equations of motion}

The equations of motion are as follows
\begin{equation}\label{aii}
\begin{split}
& d \star F_1  =  \star \tilde F_3 \wedge H_3,\cr  & d\star \tilde
F_3  =  \tilde  F_5\wedge H_3, \cr & d \star \tilde F_5  =  - F_3
\wedge H_3,
\end{split}
\end{equation}
from the R-R fields, and
\begin{equation}\label{aiii}
\begin{split}
& R-4 (\partial \pb )^2 + 4 \nabla^2 \pb -{1\over 2} \mid H_3
\mid^2=0, \cr
& d \left(e^{-2 \pb} \star H_3 \right)= F_1 \wedge
\star \tilde F_3 - \tilde F_5 \wedge \tilde F_3,
\end{split}
\end{equation}
in the NS-NS sector. The variation of the action with respect to the metric leads to
\begin{equation}
G_{MN}+e^{2 \pb}\left(g_{MN}\nabla^2 e^{-2\pb}-\nabla_M \nabla_N
e^{-2 \pb}\right)= - {2\kappa^2 \over \sqrt{-g}} {\d S_{\rm tensor}
\over \d g^{MN}}e^{2 \pb},
\end{equation}
where $G_{MN}$ is the Einstein tensor and $S_{\rm tensor}$ is the action for all the tensor fields including the dilaton. The left hand side arises from the variation of the Einstein-Hilbert action with the dilaton contribution arising from the non-canonically normalized curvature term.

Moreover, the tensor fields satisfy the Bianchi identities
\begin{equation}
\begin{split}
d H_3 & =0,\cr
d F_1 & =0,\cr
d \tilde F_3 & =  H_3\wedge F_1, \cr
d \tilde F_5 & = H_3\wedge F_3.\cr
\end{split}
\end{equation}

\subsection{The SUGRA background}

We are interested in a solution of the 10d equations of motion in
which the space-time contains four non-compact dimensions and six
compact dimension. We require maximal symmetry in the non-compact
dimensions which means all tensor fields except $F_5$ have
components along the internal directions only, while $F_5$  is
required to take the form
\begin{equation}\label{ai}
\tilde F_5 = (1+\star) d \alpha(y) \wedge dx^0 \wedge dx^1 \wedge dx^2 \wedge dx^3,
\end{equation}
where $x,y$ denote the 4d and 6d coordinates respectively.

Moreover, we would like to consider a background which arises as the
orientifold limit of a flux background of M-theory compactified on
K3$\times$K3. In this case the RR axion vanishes and the type IIB
dilaton $\pb$ is constant. The space-time metric is of the form
\begin{equation}\label{axx}
 ds^2 =
e^{2 A(y)+\pb/2}  \eta_{\mu \nu} dx^\m dx^\n  + e^{-2A(y)+\pb/2}
\left( g_{ij} dy^i dy^j + dw_1^2 + dw_2^2\right),
\end{equation}
where $g_{ij}$ is the metric of K3 and the factor involving the
dilaton arises since this is the metric in the 10d string frame and
$e^{-2A(y)}$ is the warp factor depending on the coordinates of the
internal space only. The function $\alpha$ in \C{ai} is related to
the warp factor according to
\begin{equation}\label{axxii}
\alpha(y) = e^{4 A(y)}.
\end{equation}
The complex three-form $G_3 = \tilde F_3 - i e^{-\pb}H_3$ is
imaginary self-dual in the internal dimensions, {\it i.e.}
\begin{equation}
\star G_3 = i G_3.
\end{equation}
Moreover, the warp factor satisfies the Poisson equation
\begin{equation}
\nabla^2 e^{-4 A(y)} +e^{-\pb} |H_3|^2=0.
\end{equation}
Away from the orientifold points this is a solution of the equations
of motion as can be explicitly verified.

Note that the three-form tensor fields $H_3$ and $\tilde F_3$ are
harmonic forms on the internal part of the space \C{axx}. It turns
out that the Hodge numbers of K3 are
\def\mm#1{\makebox[10pt]{$#1$}}
\begin{equation}
  {\arraycolsep=2pt
  \begin{array}{*{5}{c}}
    &&\mm{h^{0,0}}&& \\ &\mm{h^{1,0}}&&\mm{h^{0,1}}& \\
    \mm{h^{2,0}}&&\mm{h^{1,1}}&&\mm{h^{0,2}} \\
    &\mm{h^{2,1}}&&\mm{h^{1,2}}& \\ &&\mm{h^{2,2}}&&
  \end{array}} \;=\;
  {\arraycolsep=2pt
  \begin{array}{*{5}{c}}
    &&\mm1&& \\ &\mm0&&\mm0& \\ \mm1&&\mm{20}&&\mm{1} \\
    &\mm0&&\mm0& \\ &&\mm1&&
  \end{array}}
\end{equation}
and in particular there are no harmonic one-forms or three-forms on
K3. As a result $H_3$ and $\tilde F_3$ have to be the product of
harmonic two-forms on K3, which we will denote by $(h_3)_i$ and
$(\tilde f_3)_i$ and a one-form in the fiber directions, $dw^i$,
{\it i.e.}
\begin{equation}
H_3 = (h_3)_i\wedge  dw^i \qquad {\rm and } \qquad \tilde F_3 = (\tilde f_3)_i \wedge dw^i, \qquad i =1,2,
\end{equation}
where $w_i\sim w_i+1$ and
\begin{equation}
(\tilde f_3)_i ,~ (h_3)_i \in H^2(K3,\IZ).
\end{equation}
Moreover, the condition that $G_3$ is imaginary self-dual
requires the complex three-form to be
\begin{equation}
G_3 = g_+ \wedge d \bar w + g_- \wedge dw,
\end{equation}
where
\begin{equation}
dw = dw_1 + i dw_2,
\end{equation}
and $g_{\pm}$ can be expanded in (anti)-self dual harmonic two-forms
on K3
\begin{equation}
g_+ \in H^{2,0}(K3)\oplus H^{0,2}(K3) \oplus H^{1,1}_+(K3)\qquad {\rm and } \qquad g_- \in H_-^{1,1}(K3).
\end{equation}
There are 3 self-dual two-forms and 19 anti-self dual two-forms
which are of type $(1,1)$ and primitive. In the following we will
see that the different solutions of the equations of motion preserve
different amounts of supersymmetry. In particular, the amount of
unbroken supersymmetry will depend on the choices of two-forms on
K3.

\subsection{Supersymmetry}

Let us represent the dilatino and gravitino fields by Weyl spinors
$\l$ and $\Psi_\m$, respectively. Similarly, the infinitesimal
supersymmetry parameter is represented by a Weyl spinor $\e$. The
supersymmetry transformations of the fermi fields of type IIB
supergravity (to leading order in fermi fields) are
\begin{equation}
\d \l = \half \left(\pa\!\!\!/ \pb -i e^{\pb} \pa\!\!\!/ C_0 \right)
\e + { 1 \o 4}\left(i e^{\pb}\widetilde {F}_3\!\!\!\!\!/~~ - {
H}_3\!\!\!\!\!\!/ ~ \right) \e^\star ,
\end{equation}
and
\begin{equation}
\d \Psi_M = \left( \nabla_M + {i\o 8} e^{\pb} { F}_1 \!\!\!\!\!/ ~~\G_M
+\frac{i}{16} e^{\pb} \widetilde{ F}_5\!\!\!\!\!/ ~~ \G_M \right)\e
-\frac{1}{8} \left(2({H}_3)_M \!\!\!\!\!\!\!\!\!\!\!\!\!/ ~~~~~  + i e^{\pb} \widetilde{
F}_3 \!\!\!\!\!/ ~~ \G_M   \right) \e^\star.
\end{equation}

Upon reducing to 4d the Lorentz algebra decomposes according to
$SO(9,1) \to SO(3,1) \times SO(6)$. The Weyl spinor $\e$ then
decomposes as
\begin{equation}
{\bf 16} \to ({\bf 2}, {\bf 4}) + ({\bf 2'}, {\bf 4}').
\end{equation}
Under the further decomposition $SO(6) \to SO(4) \times SO(2) $
\begin{equation}
\begin{split}\label{bbi}
{\bf 4}& \to  ({\bf 2} , {\bf 1}) + ({\bf 2'},{\bf 1'})\cr
{\bf 4'}& \to  ({\bf 2} , {\bf 1'}) + ({\bf 2'}, {\bf 1})\cr
\end{split}
\end{equation}
The holonomy of K3 is $SU(2)$ and under the reduction $SO(4) \to
SU(2)$
\begin{equation}
\begin{split}
{\bf 2 } & \to   {\bf 1} + {\bf 1}\cr
{\bf 2'} & \to  {\bf 2}.\cr
\end{split}
\end{equation}
This means that either ${\bf 4}$ or ${\bf 4'}$ of $SO(6)$ gives rise
to two $SU(2)$ singlets leading to an N=4 supersymmetry in 4d.

Next we analyze the constraints imposed by the orientifold projection $\IZ_2 = \Omega (-1)^{F_L} {\cal I}$.
Writing $\e=\e_1+i \e_2$ the different parity transformations act according to
\begin{equation}
\e=\e_1+i \e_2 \xrightarrow{\Omega} \e_2+i
\e_1\xrightarrow{(-1)^{F_L}} -\e_2+i \e_1\xrightarrow{\cal I } i
\Gamma_{\star} (-\e_2 + i \e_1),
\end{equation}
where $\Gamma_\star$ is the chirality operator of $SO(2)$. Combining these operations and requiring
the spinor to be left invariant by the orientifold action imposes
\begin{equation}
\e=-\Gamma_\star \e.\label{spinor}
\end{equation}

Before we proceed, lets determine how the spinor projection relates
to the one in the type I string. After two T-dualities on torus, the
left moving spinor $\e_1$ is unaffected, however the right moving
spinor $\e_2$, transforms as
\begin{equation}
\e_2\to \G^8\G^9\e_2,
\end{equation}
from which we get the transformation of Eq.(\ref{spinor}),
\begin{equation}
(1+\G_\star)(\e_1-\e_2)=0
\end{equation}
Because the gamma matrix $\G_\star$ is pure imaginary in our
representation, this condition leads to $\e_1=\e_2$, the spinor that
survives the world sheet projection of type IIB string, {\it i.e.}
type I string. This is an alternative way to see how type I string
emerges after performing T-dualities of type IIB orientifold.

Eqn.(\ref{spinor}) means that spinor has a definite chirality on the
torus, which we choose to be ${\bf 1}$ in eqn. \C{bbi}, while ${\bf
1'}$ is projected out. As a result the $SU(2)$ singlets which are
not projected out by the orientifold arise from the ${\bf 4}$ in
eqn. \C{bbi}. The orientifold breaks the 4d supersymmetry from N=4
to N=2. Moreover, the two 4d spinors are in the $\bf 2$ of $SO(3,1)$
so have the same chirality. We denote the resulting spinors by
$\eta_i$, and by an $SO(4)$ transformation we can choose them to
satisfy
\begin{equation}\label{bbiii}
\Gamma_i \eta_1 = \Gamma_w \eta_1=0 \qquad {\rm and } \qquad \Gamma_{\bar i} \eta_2 = \Gamma_w \eta_2=0,
\end{equation}
where $(y^i,y^{\bar i})$ and $(w,\bar w)$ are complex coordinates on
K3 and the torus respectively.

Using these supersymmetry transformations the unbroken
supersymmetries are those that satisfy $\delta({\rm fermi})=0$.
Evaluated in the background metric \C{axx}, using the relation
between the warp factor $A(y)$ and $\alpha(y)$ and the fact that the
spinors have definite 4d chirality the supersymmetry conditions
become
\begin{equation}
\nabla_i \left( e^{-A/2} \e\right) =0,
\end{equation}
which is satisfied with a spinor proportional to the covariantly
constant spinors on K3$\times$T$^2$ and
\begin{equation}\label{bbii}
{\bf G}_m \e^\star=0 \qquad {\rm and } \qquad \bf G \e =0.
\end{equation}

Next we solve the constraints \C{bbii} and we will check that
depending on the choice of flux different amounts of supersymmetry
are preserved. Lets analyze the amount of unbroken supersymmetry

\begin{itemize}

\item[\tria] if $G=g_- \wedge dw$,  then
\begin{equation}
G_{w i \bar j} \Gamma^{i \bar j} \eta_k^\star = G_{i w \bar j} \Gamma^{w \bar j} \eta_k^\star
=G_{\bar j i w } \Gamma^{i w} \eta_k^\star = G_{w i \bar j } \Gamma^{ w i \bar j } \eta_k=0,
\end{equation}
for $k=1,2$. This is solved by requiring $G$ to be primitive with
respect to the base, {\it i.e.}
\begin{equation}\label{di}
G_{w i \bar j} g^{i \bar j}=0,
\end{equation}
while both spinors $\eta_k$ for $k=1,2$ are non-vanishing. Since
$g_-$ are expanded in a basis of anti-self dual (1,1) forms eqn.
\C{di} is always satisfied. This leads to an N=2 supersymmetry in
4d.

\item[\tria] if $G=g_+^{2,0}\wedge d \bar w$, eqn. \C{bbii} requires
\begin{equation}
G_{{\bar w} ij} \Gamma^{ij}\eta_2^\star=0,
\end{equation}
which is solved by $\eta_2=0$, while the conditions on $\eta_1$ are
\begin{equation}
G_{{\bar w} ij}\Gamma^{ij} \eta_1^\star = G_{i \bar w j} \Gamma^{\bar w j} \eta_1^\star
= G_{i j \bar w} \Gamma^{ij \bar w} \eta_1.
\end{equation}
These conditions are always satisfied which implies that the 4d
supersymmetry arising from $\eta_1$ is unbroken. This flux
configuration leads to an N=1 supersymmetry in 4d.

\item[\tria] if $G=g_+^{0,2}\wedge d \bar w$, eqn. \C{bbii} requires
\begin{equation}
G_{\bar w \bar i \bar j} \Gamma^{\bar i \bar j} \eta_k^\star = G_{\bar i\bar
w \bar j} \Gamma^{\bar w \bar j} \eta_k^\star
=G_{\bar j \bar i\bar w } \Gamma^{\bar i \bar w} \eta_k^\star = G_{\bar w \bar
i \bar j } \Gamma^{ \bar w \bar i \bar j } \eta_k=0,
\end{equation}
for $k=1,2$. These conditions are solved by requiring $\eta_1=0$
while $\eta_2 \neq 0$ and as a result there is an N=1' unbroken
supersymmetry in 4d. We label this supersymmetry with N=1' since it
preserves a different subgroup of the supersymmetry than the
$G_{{\bar w } ij}$ component.

\item[\tria] if $G=g_+^{1,1}\wedge d \bar w$, eqn. \C{bbii} requires $\eta_1=\eta_2=0$ and
supersymmetry is completely broken.

\end{itemize}

\section{Heterotic SUGRA background}

In this section we analyze the heterotic flux backgrounds. To set up
the notation we review the heterotic low-energy effective action to
$O(\alpha'^2)$ in section 3.1. In section 3.2 we summarize the
equations of motion. In section 3.3 we present the backgrounds
solving the SUGRA equations to leading order in $\alpha'$. In
section 3.4 we analyze the amount of unbroken four-dimensional
supersymmetry. This section is confined to solutions solving the
SUGRA equations to leading order in  $\alpha'$ and the corrected
background is discussed in section 4.

\subsection{The action}
The bosonic part of the heterotic supergravity action to $O(\a'^2)$
in the 10d string frame is
\cite{Callan:1985ia}\cite{Metsaev:1987zx}\cite{Bergshoeff:1988nn}\cite{Bergshoeff:1989de}\cite{Chemissany:2007he}
\begin{equation}\label{ri}
S_{het}={1\o 2\kappa^2}\int d^{10}x\sqrt{-g} e^{-2\ph} \left[R+4
(\p\ph)^2 -{1\o2}|{\cal H}|^2-{\alpha'\over 4} {\rm tr} ({\cal
F}^2-R_+^2)\right],
\end{equation}
where
\begin{equation}
{\cal H} =d{\cal B} +{\a'\o4}\omega_3,
\end{equation}
is the NS three-form and ${\cal F}=d{\cal A} + {\cal A}\wedge {\cal
A}$ is the gauge field strength. Moreover, $\omega_3=\omega_{\rm
L}-\omega_{\rm YM}$ is given in terms of the Lorentz and Yang-Mills
Chern-Simons three-forms
\begin{equation}
\omega_{\rm L} = {\rm tr} \left(\O_+ \wedge d\O_+ +{2\over 3} \O_+
\wedge \O_+ \wedge \O_+ \right) \quad {\rm and } \quad \omega_{\rm
YM} = {\rm tr} \left({\cal A}\wedge d{\cal A} +{2\over 3}{\cal A}
\wedge {\cal A} \wedge {\cal A} \right).
\end{equation}
The contribution to the action which is quadratic in the Riemann
tensor is
\begin{equation}
{\rm tr} R_+^2 = {1\o2}R_{MNAB}(\O_+)R^{MNAB}(\O_+),
\end{equation}
while the quadratic term in ${\cal F}$ is the standard gauge field
kinetic term. Note that the Einstein-Hilbert action is formulated in
terms of the spin connection while the quadratic term in the Riemann
tensor is expressed in terms of a connection involving the NS
three-form which explicitly is defined by
\begin{equation}
{\Omega_{\pm}^{AB}}_M ={\Omega^{AB}}_M\pm {1\over 2} {{\cal
H}^{AB}}_M.
\end{equation}
Also, we will follow ref. \cite{Bergshoeff:1989de} according to
which the action involves the $\Omega_+$ connection while the
supersymmetry transformations involve the $\Omega_-$ connection. The
supersymmetry tranformations will be described in more detail below.

\subsection{Equations of motion}

The equations of motion arising from the action presented in the
previous section are

\begin{itemize}

\item[\tria] for the dilaton
\begin{equation}
R-4(\nabla\ph)^2+4\nabla^2\ph-{1\o2}|{\cal H} |^2-{\alpha'\over 4}
{\rm tr} ({\cal F}^2-R_+^2)=0,\label{het_dil_eom}
\end{equation}

\item[\tria] for ${\cal B}$
\begin{equation}
d(e^{-2\ph}\star_{10} {\cal H} )=0,\label{b2}
\end{equation}

\item[\tria] for the metric
\begin{eqnarray}
\begin{split}
& R_{MN}+2\nabla_M\nabla_N \ph-{1\o 4}{\cal H}_{MPQ}{\cal
H}_{N}{}^{PQ} + \cr & {\a'\o4}[R_{MPQR}(\O_+)R_N{}^{PQR}(\O_+)-{\cal
F}_{MP}{\cal F}{}_N{}^P]=0, \end{split} \label{het_g_eom}
\end{eqnarray}

\item[\tria] for the Yang-Mills field
\begin{equation}
e^{2\ph}d(e^{-2\ph}\star_{10}{\cal  F} )+{\cal A}\wedge
\star_{10}{\cal F}-\star_{10}{\cal F}\wedge {\cal A} +{\cal
F}\wedge\star_{10} {\cal H}=0. \label{het_a_eom}
\end{equation}

\end{itemize}

The Bianchi identities are
\begin{equation}\label{bianchi}
d{\cal H} ={\alpha'\over 4} [{\rm tr} (R_+\wedge R_+)-\tr( {\cal
F}\wedge {\cal F})] \qquad {\rm and } \qquad d{\cal F}+[{\cal
A},{\cal F}] = 0.
\end{equation}

\subsection{The SUGRA background}

In the following, we present the background that solves the SUGRA
equations of motion to leading order in $\alpha'$ (see
ref.\cite{DRS:M-theory}\cite{Becker:2002sx}\cite{Becker:2006et} for
supersymmetric backgrounds). As we will see non-trivial solutions of
the Bianchi identity exist only for non-compact backgrounds. This
conclusion is modified once $\alpha'$ corrections are taken into
account.

 The background metric is
\begin{equation}
ds_{het}^2=\eta_{\m\n}dx^\m dx^\n+e^{-4A(y)}
g_{ij}dy^idy^j+E_{w_1}E_{w_1}+E_{w_2}E_{w_2},\label{metric}
\end{equation}
where
\begin{equation}
E_{w_1}=dw_1+B_{iw_1}dy^i\qquad {\rm and } \qquad
E_{w_2}=dw_2+B_{iw_2}dy^i,
\end{equation}
and $B_{w_k}=B_{i w_k} dy^i$, for $k=1,2$ are one-forms on the base.
These one-forms are constrained by the condition that
\begin{equation}\label{h_2}
H_{w_1} = dB_{j w_1}dy^j\qquad {\rm and } \qquad
H_{w_2} = dB_{j w_2}dy^j,
\end{equation}
are harmonic non-trivial two-forms on K3. Note that $E_{w_k}$ have
to be globally defined since otherwise the metric is not be globally
defined. As a result on the 6d space $H_{w_k}=dE_{w_k}$ become exact
even though these forms are non-trivial on K3. We will expand
$H_{w_k}$ in harmonic non-trivial two-forms on K3. Depending on the
choice of flux different amounts of 4d supersymmetry will preserved
as we will see in the next section.

The three-form is
\begin{equation}\label{dddi}
{\cal H}= e^{2 \ph} \star_6 d \left(e^{-2 \ph} E^{w_1} \wedge
E^{w_2}
 \right)
 =\star_b d  e^{-4 A(y)}- \star_b H_{w_1} \wedge E^{w_1} - \star_b H_{w_2} \wedge E^{w_2},
\end{equation}
where $\star_6$ denotes the Hodge dual with respect to the 6d
internal space and $\star_b$ denotes the Hodge dual with respect to
the unwarped base.

The dilaton is given by
\begin{equation}
\ph=-2 A(y).
\end{equation}

The Yang-Mills field is assumed to be a two-form on K3 only and to
satisfy the hermitian Yang-Mills equations, {\it i.e.}
\begin{equation}
{\cal F}_{i \bar j} J^{i \bar j} =0 \qquad {\rm and  } \qquad {\cal
F}_{ij}={\cal F}_{\bar i \bar j}=0.
\end{equation}
Here $J$ is the K\"ahler form of K3.

Moreover, $A(y)$ is a scalar function depending on the coordinates
of the base only. To leading order it is required to solve the
differential equation
\begin{equation}\label{zzxi}
\nabla^2 e^{-4 A(y) } + \mid H_{w_1} \mid^2+\mid H_{w_2} \mid^2=0.
\end{equation}

Next we show that this background satisfies the equations of motion
to leading order in $\alpha'$. The equation of motion of ${\cal B}$
is satisfied since \C{dddi} implies
\begin{equation}\label{dddii}
\star_{10} {\cal H} = -e^{ 2 \ph} d\left(e^{-2 \ph} E^{w_1} \wedge E^{w_2}\right) \wedge dx^{0123}.
\end{equation}

The equation of motion for the metric has several components
\begin{equation}
(\m,\n), ~(i,j),~(w_1, i), ~(w_2,i),~(w_1,w_2).
\end{equation}
The $(i,j)$ component, with two indices on K3, is satisfied assuming
$A(y)$ solves \C{zzxi}. Moreover, it is easy to see that all other
components vanish to this order in $\alpha'$.

Next we consider the dilaton equation of motion. Using the metric
\C{metric} to compute the scalar curvature $R$, the dilaton equation
of motion is solved assuming $A(y)$ solves eqn. \C{zzxi}. On the
other hand the Bianchi identity leads to
\begin{equation} \label{dh}
d {\cal H}=  -\left(\nabla^2 e^{-4 A(y)} +  \mid H_{w_1} \mid^2+\mid
H_{w_2}\mid^2 \right)\star_b 1= 0,
\end{equation}
which again is solved after imposing eqn. \C{zzxi}

Note that eqn. \C{zzxi} only has non-trivial solutions if the
internal space is non-compact. Below we will describe in detail how
to construct compact solutions by going beyond the leading order in
$\alpha'$.

\subsection{Supersymmetry}
Next let us analyze the supersymmetry of the solutions of the
equation of motion. The supersymmetry transformations leaving the
10d heterotic string frame effective action invariant are
\begin{equation}
\begin{split}
& \d \Psi_M  =  \nabla_M \ve - \fourth {{\cal H}_{M}}\!\!\!\!\!\!\!\!\!/ ~~~\ve, \nonumber \\
 & \d \l  =  \paslash \phi_h \ve - \half {\cal H}\!\!\!\!/ ~\ve,\\
 & \d \c  =  2 {\cal F}\!\!\!\!/ ~\ve \nonumber,
\end{split}
\end{equation}
where $\Psi_M$ is the gravitino, $\l$ the dilatino and $\c$ the gaugino. All spinors are Majorana-Weyl.
The covariant derivative of a spinor is defined according to
\begin{equation}
\nabla_M \epsilon = \partial_M\ve+\fourth {\Omega^{AB}}_M \Gamma_{AB} \ve,
\end{equation}
where $\Omega$ is the spin connection. Note that the gravitino
variation can then be written in the form
\begin{equation}
\d \Psi_M  =  \partial_M \ve + \fourth
{\Omega_{-}^{AB}}_M\Gamma_{AB}\ve,
\end{equation}
where
\begin{equation}\label{kkx}
{\Omega_{\pm}^{AB}}_M={\Omega_{}^{AB}}_M\pm \half {H_{}^{AB}}_M.
\end{equation}

Explicitly the components of the spin connection are
\begin{equation}
\begin{split}
{\O_\pm^{w_1 }}_a  = & \half e^{2 A} {\left(H_{w_1} \mp \star_b
H_{w_1} \right)}_{ab}e^b , \cr {\O_\pm^{w_2 }}_a  = & \half e^{2
A}{\left(H_{w_2} \mp \star_b H_{w_2}  \right)}_{ab}e^b , \cr
{\O_\pm^{a}}_{b}  = & 2\left[\partial^a A e_b -
\partial_b A e^a \mp  {(\star_b d A)^
{a}}_{bc}e^c\right] +{\omega^{a}}_b \cr & -\half e^{4
A}{\left(H_{w_1} \pm \star_b H_{w_1} \right)^{a}}_bE^{w_1}-\half
e^{4 A} {\left(H_{w_2} \pm \star_b H_{w_2} \right)^{a}}_bE^{w_2}.
\cr
\end{split}
\end{equation}
Note the sign differences between the first two components of the
spin connection and the last one. These sign differences will play a
crucial role in the supersymmetry analysis.

Under the decomposition $SO(9,1) \to SO(3,1)\times SO(6)$ a 10d Weyl
spinor decomposes as ${\bf 16} \to ({\bf 2}, {\bf 4})+ ({\bf
2}',{\bf 4}')$. Imposing the Majorana condition we set
\begin{equation}
\epsilon = \zeta \otimes \eta + \zeta^\star  \otimes \eta^\star,
\end{equation}
where $\zeta \otimes \eta$ transforms as $({\bf 2}, {\bf 4})$. Since
the complex conjugate is not an independent spinor each 6d Weyl
spinor gives rise to one minimal 4d supersymmetry.

Lets solve the supersymmetry constraints. The gravitino condition
with the index in the external space-time is satisfied if the spinor
does not depend on the coordinates of the external space-time.
Projecting onto spinors with definite 4d chirality the supersymmetry
conditions become
\begin{equation}
\begin{split}
& \nabla_M \eta - \fourth {{\cal H}_{M}}\!\!\!\!\!\!\!\!\!/ ~~~\eta=0, \nonumber \\
& \paslash \ph \eta - \half {\cal H}\!\!\!\!/ ~\eta=0,\\
&{\cal F}\!\!\!\!/ ~\eta=0 \nonumber,
\end{split}
\end{equation}
which are equations constraining the 6d spinor $\eta$. To solve this
supersymmetry conditions the spinor $\eta$ has to satisfy
\begin{equation}
\partial_{w_i}\eta=0\qquad {\rm and } \qquad \partial_i \eta + {1\over 4} {\omega^{ab}}_i\gamma_{ab} \eta=0,
\end{equation}
{\it i.e.} $\eta$ is a covariantly constant spinor on the base. We
denote the two covariantly constant spinors of K3 by $\eta_k$,
$k=1,2$. Moreover, we require $\eta_{k}$ to solve
\begin{equation}
\begin{split}
& (H_{w_1}-\star_b H_{w_1})_{ab} \gamma^{ab} \eta_{k}=
(H_{w_2}-\star_b H_{w_2})_{ab} \gamma^{ab} \eta_{k}=0,\cr &
{(H_{w_1}+\star_b H_{w_1})_{ab}} \gamma^{w_1a} \eta_{k} +
{(H_{w_2}+\star_b H_{w_2})_{ab}} \gamma^{w_2a} \eta_{k}=0,
\end{split}
\end{equation}
which after introducing complex coordinates $w=w_1+i w_2$, so that
\begin{equation}
H_{w}=\half( H_{w_1}-i H_{w_2})\qquad {\rm and } \qquad  H_{\bar
w}=\half( H_{w_1}+i H_{w_2}),
\end{equation}
take the form
\begin{equation}\label{zzi}
\begin{split}
& \left[ (1-\star_b) {H_w}\right]_{ab}\gamma^{ab}\eta_{k}=0, \cr &
\left[ (1-\star_b) {H_{\bar w} }\right]_{ab} \gamma^{ab}\eta_{k}=0,
\cr & {[(1+\star_b)H_w]_{ab}}{\gamma}^{wa}\eta_k +
{[(1+\star_b)H_{\bar w}]_{ab}}\gamma^{{\bar w}a}\eta_k=0.
\end{split}
\end{equation}

Note that the contributions involving the warp factor arising from the spin connection components
${\O_\pm^{ab}}_{c} $ and contributing to the component of the gravitino variation along the base cancel since the two spinors $\eta_{k}$ have positive chirality on the base {\it i.e.}
\begin{equation}
-\gamma_{1234} \eta_k = \eta_k\qquad k=1,2.
\end{equation}

Now depending on the choice of flux different amounts of
supersymmetry are preserved \cite{Becker:2006et}. The different
cases are
\begin{itemize}
\item[\tria] if $H_w$ is
proportional to an anti-self dual (1,1) form on the K3 base, the
conditions \C{zzi} are satisfied for both spinors $\eta_k$, $k=1,2$.
An N=2 supersymmetry is preserved in 4d. Indeed, the third condition
is trivially satisfied and the first two conditions are satisfied
since the anti-self dual (1,1) forms are primitive with respect to
the base.

\item[\tria] if $H_w$ is proportional to the self-dual (0,2) form on the
base the supersymmetry generated by $\eta_1$ is preserved while
$\eta_2=0$. There is an N=1 unbroken supersymmetry in 4d.

\item[\tria] if $H_w$ is proportional to the self-dual (2,0) form on the
base the supersymmetry generated by $\eta_2$ is unbroken while
$\eta_1=0$. There is an N=1' unbroken supersymmetry in 4d.

\item[\tria] if $H_w$ is proportional to the self-dual (1,1) form on
the base \C{zzi} requires the two spinors to vanish. So N=0 in 4d.

\end{itemize}

\section{The $\alpha'$ corrected torsional heterotic geometry}

In this section we will consider $\alpha'$ corrections to the
torsional heterotic geometries. We will see that these $\alpha'$
corrections to the background are required since otherwise the
$\alpha'$ corrected equations of motion are not satisfied. Once the
background is corrected in $\alpha'$ compact solutions become
possible. As a first step to solve the Bianchi identity we need to
compute tr$(R_+\wedge R_+)$, which appears on the right hand side of
the Bianchi identity.

\subsection{tr$(R_+\wedge R_+)$}

In general, the curvature two-form is defined by
\begin{equation}
{R^A}_B={d\Omega^A}_B+{\Omega^A}_C\wedge {\Omega^C}_B,
\end{equation}
for some connection $\Omega$. According to Bergshoeff and de Roo
\cite{Bergshoeff:1989de} the connection used in the supersymmtry
transformations is $\Omega_-$ while in the Bianchi identity the
$\Omega_+$ connection is used. The connection coefficients are
\begin{equation}\label{zzii}
\begin{split}
{\O_+^{w_k }}_a  = & \half e^{2 A} {\left(H_{w_k} - \star_b H_{w_k}
\right)}_{ij}e^i_a dy^j ,\qquad k=1,2 \cr {\O_+^{a}}_{b}~ = &
{\s^a}_b +{\omega^{a}}_b -\half {\left(H_{w_k} + \star_b H_{w_k}
\right)}_{ij} E^i_c E^j_b \eta^{ac} E^{w_k},
\end{split}
\end{equation}
where, the last term involves a sum over $k=1,2$. We denote with
$E^a$ the vielbeine of the warped base while $e^a$ are those of the
unwarped K3. Moreover,
\begin{equation}
\s_{ab}= 2\left[ \partial_a A e_b -
\partial_b A e_a - (\star_b d A)_{abc}e^c\right].
\end{equation}
Note that $\s_{ab}$ is self-dual in its indices, {\it i.e.} it
satisfies
\begin{equation}
\s_{ab} = \half \e_{abcd} \s^{cd}.
\end{equation}
We are denoting the spin connection coefficients and curvature
two-form of the K3 base by ${\omega^a}_b$ and ${r^a}_b$.

Before describing in detail the results for the curvature two-form
and $\tr (R_+\wedge R_+)$, where $R_+$ is computed with respect to
the $\Omega_+$ connection, we will first establish that the
curvature two-form of the torsional space is of type (1,1) to
leading order in $\a'$ if computed with respect to the $\Omega_+$
connection. This implies that $\tr (R_+\wedge R_+)$ is a (2,2) form
which is a necessary condition for the Bianchi identity to admit a
non-trivial solution. Indeed, up to terms of $O(\a'^2)$ unbroken
supersymmetry requires the flux and the fundamental (1,1) form to be
related according to $ {\cal H} = i (\partial- \bar
\partial ) J$. As a result $d {\cal H}= - 2 i \partial \bar \partial
J$ is a (2,2) form. This is the left hand side of the Bianchi
identity. The right hand side of the Bianchi identity is $\tr
(R_+\wedge R_+)$, which is required to be a four-form of type (2,2)
since otherwise the background is over-constrained.

Here we follow the presentation of ref. \cite{Sen:1986mg}. By
definition
\begin{equation}\label{kkx}
{\Omega_{+}^{AB}}_{M}={\Omega_{}^{AB}}_{M}+ \half {H_{}^{AB}}_{M},
\end{equation}
which implies that the connection in the coordinate basis is
modified to
\begin{equation}
\G_{+IK}^J=G^{JL}(E_L^A\partial_I E^A_K + {\Omega_+^{AB}}_I E_L^A
E_K^B) = \Gamma_{IK}^J - \half {{\cal H}_{IK}}^J.
\end{equation}
By definition
\begin{equation}\label{kkxi}
\G_{IK}^J=\half g^{JN} \left(\pa_I g_{NK} + \pa_K g_{NI}-\pa_N
g_{IK} \right) .
\end{equation}

Supersymmtry requires ${\cal H}$ to be related to the derivative of
the metric according to
\begin{equation}\label{kkxii}
{\cal H}_{ M N \bar P} =-\partial_{ M} g_{N \bar P}+\partial_{ N}
g_{ M \bar P},
\end{equation}
and the complex conjugate. Here we have introduced complex
coordinates. Using the fact that the metric of the torsional space
is hermitian eqn. \C{kkxi} implies that the non-vanishing connection
coefficients are
\begin{equation}
\G^I_{+ JK}=g^{I \bar N} \partial_J g_{K \bar N} \qquad {\rm and }
\qquad \G^I_{+J \bar K} =g^{I \bar N} \pa_{\bar K} g_{J \bar N } -
g^{I \bar N} \partial_{\bar N} g_{J \bar K}.
\end{equation}
So in contrast to K\"ahler geometry there are connection
coefficients with mixed indices.

The Riemann tensor is obtained from the connection coefficients
according to
\begin{equation}
R_{MN}{}^{K}{}_L=\pa_M \G_{NL}^K - \partial_N \G_{ML}^K + \G_{MR}^K
\G_{NL}^R - \G_{NR}^K \G_{ML}^R,
\end{equation}
and the curvature two-form is related to the Riemann tensor
according to
\begin{equation}
{R^A}_B=\half {R_{CD}{}^A{}_B} E^C E^D.
\end{equation}

Introducing complex coordinates it is not difficult to see that
\begin{equation}
R_{+MN}{}^{ K}{}_L=R_{+M N}{}^{\bar K}{}_L=R_{+ MN}{}^{\bar
K}{}_{\bar L}=0.
\end{equation}
Moreover,
\begin{equation}
R_{+\bar M\bar N}{}^{ \bar K}{}_{L} =g^{P\bar K} \left(g_{P [ \bar
N,\bar M] L} - g_{ L [ \bar N,\bar M] P} \right) =O(\alpha').
\end{equation}
This quantity is subleading since the right hand side is the (2,2)
component of $d{\cal H}$ which is $ O(\alpha')$ after using the
Bianchi identity. Therefore we conclude that to leading order in
$\alpha'$, $\tr (R_+ \wedge R_+)$ is of type (2,2).

Next we present the explicit results for the curvature two-forms and
$\tr (R_+ \wedge R_+)$ and show how to solve the Bianchi identity.
We will focus on solutions with N=2 supersymmetry.

\subsection{N=2 background at $O(\a')$}

In this case the forms $H_{w_i}$ are proportional to anti-self dual
(1,1) forms on the K3 base. From \C{zzii} we see that the only
non-vanishing components of the spin connection are
\begin{equation}
\begin{split}
{\O_+^{w_k }}_a  = &  e^{2 A} {\left(H_{w_k} \right)}_{ij}e^i_a dy^j
, \qquad k=1,2 \cr  {\O_+^{a}}_{b}~ = & {\s^a}_b+{\omega^{a}}_b .
\end{split}
\end{equation}

In this case the curvature two-form computed with respect to the
$\Omega_+$ connection is a two-from on K3 explicitly given by
\begin{equation}
\begin{split}
{R^{w_1}}_{w_2}& =-e^{4 A} (H_{w_1})_{a} {(H_{w_2})^a}  \cr
{R^a}_{w_k}~& =-\nabla[e^{2 A} (H_{w_k})_a]-e^{2A} (H_{w_k})_b
{\s^b}_a, \qquad k = 1,2 \cr {R^a}_b~~~& ={r^a}_b+\nabla
{\s^a}_c+{\s^a}_c {\s^c}_b -e^{4 A} (H_{w_k})_a (H_{w_k})_b , \cr
\end{split}\label{n2}
\end{equation}
where ${r^a}_b$ is the curvature two-form of K3 and $\nabla$ is the
covariant derivative with respect to the ${\omega^a}_b$ connection.
Explicitly
\begin{equation}
\nabla {\s^a}_b=d{\s^a}_b+{\omega^a}_c  {\sigma^c}_b+{\sigma^a}_c
{\omega^c}_b.
\end{equation}

A convenient way to compute $\tr (R_+\wedge R_+)$ is to use the
Chern-Simons formula which relates the results for $\tr (R\wedge R)$
computed with two connections $\G$ and $\tilde \G$ according to
\begin{equation}
\tr (R\wedge R) - \tr (\tilde R \wedge \tilde R )= d Q(\G, \tilde
\G),
\end{equation}
where
\begin{equation}
Q(\G, \tilde \G)= 2 \alpha \wedge R - \alpha \wedge d \alpha - 2
\alpha \wedge \G \wedge \alpha +{2 \over 3} \alpha\wedge \alpha
\wedge \alpha
\end{equation}
where $\alpha=\G - \tilde \G$. Setting
\begin{equation}
\begin{split}
&  {{{\tilde \G}^a}}_{~b}={\Omega_+^a}_b \qquad {\rm and } \qquad {
{{ \tilde \G} ^{w_k}} }_{~~a}= { {{ \Omega_+} ^{w_k}} }_{a}\cr &
{\G^a}_b={\Omega_+^a}_b\qquad {\rm and } \qquad { {{ \G} ^{w_k}}
}_{a}=0, \qquad k=1,2,\cr
\end{split}
\end{equation}
or in other words choosing
\begin{equation}
{\a^a}_b=0\qquad {\rm and } \qquad {\a^{w_k}}_a = - e^{2 A}
(H_{w_k})_{ij}e^i_a dy^j ,
\end{equation}
we obtain
\begin{equation}
\tr (R_+ \wedge R_+)  =   \tr [R(\G) \wedge R(\G)]
 +2  d \left\{e^{ 2 A} (H_{w_k})_b \nabla [e^{2 A} (H_{w_k})_b]
 +e^{4 A} (H_{w_k})_b {\s^b}_c  (H_{w_k})_c \right\},
 \end{equation}
where
\begin{equation}
\tr [R(\G) \wedge R(\G)] = -(\nabla {\s^a}_b +{r^a}_b+{\s^a}_c
{\s^c}_b ) (\nabla {\s^b}_a +{r^b}_a+{\s^b}_c {\s^c}_a )
\end{equation}
This result can be further simplified by using the Chern-Simons
formula again, this time with
\begin{equation}
{\tilde \G^a}_{~b}={\omega^a}_b \qquad {\rm and } \qquad
{\G^a}_b={\omega^a}_b+{\sigma^a}_b .
\end{equation}
The result is
\begin{equation}
\tr [R(\G) \wedge R(\G)]=\tr ( r \wedge r) -2^4 d\left[2 (\nabla^2
A)\star d A - \star d  (\nabla A)^2 - 8 (\nabla A)^2 \star dA
\right].
\end{equation}

A straightforward but tedious computation then shows
\begin{equation}\label{rr1}
\begin{split}
\tr (R_+ \wedge R_+ )  =  & \tr( r \wedge r) +4 d \star_b d
\left(\nabla^2 A \right)+ \cr &  d \star_b d \left[(\nabla^2 e^{-4A}
+|H|^2)e^{4A} \right]+ \cr & 2 d \left[(\nabla^2 e^{-4A}
+|H|^2)\star_b d e^{4A} \right],
\end{split}
\end{equation}
where
\begin{equation}
|H|^2=|H_{w_1}|^2+|H_{w_2}|^2.
\end{equation}
Note that the last two lines in eqn. \C{rr1} involve the leading
order equation of motion \C{zzxi}. Thus we establish that for
solutions preserving an N=2 supersymmetry in four dimensions $\tr
(R_+ \wedge R_+)$ is a (2,2) form with components along the K3 base
only. Note that this fact is a consequence of having used the
$\Omega_+$ connection to compute $\tr (R_+ \wedge R_+)$. Since $\tr
(R_+ \wedge R_+)$ has components along the base only the fiber is
not required to be of $O(\a')$ and can be chosen to be large.

Next we will use this result and solve the Bianchi identity
\begin{equation}
d{\cal H}={\a'\o 4}[\tr (R\wedge R)-\tr ({\cal F}\wedge{\cal  F})]
\end{equation}
to $O(\a')$.

First we note that the second and third line on the right hand side
of Eq.(\ref{rr1}) are proportional to the dual of $d{\cal H}$ and
are therefore $O(\a')$.  As a result they contribute to the Bianchi
identity only to $O(\a'^2)$. Keeping all terms up to $O(\a')$ the
Bianchi identity becomes
\begin{equation}\label{vi}
d \star_b d e^{-4 A} - \star_b H_{w_k} \wedge H_{w_k}+O(\a') = {
\a'\over 4} [\tr( r \wedge r)-\tr ({\cal F} \wedge {\cal F})] +\a' d
\star_b d (\nabla^2 A).
\end{equation}
Here we have allowed a correction to $O(\a')$ on the left hand side.
Since the supersymmtry transformations receive only corrections at
$O(\a'^2)$ any corrections to the left hand side of eqn.(\ref{vi})
have to solve the leading order supersymmetry conditions. Since the
supersymmtry conditions do not determine $A(y)$ we can redefine the
warp factor and still obtain a supersymmetric situation. In
particular if we define
\begin{equation}
e^{-4A'}=e^{-4A}+\a' \nabla^2 A.
\end{equation}
and allow the background to receive an $O(\a')$ correction according
to
\begin{equation}
\begin{split}\label{corrected}
& \ph=-2A'(y),\cr & {\cal H}=\star_b d  e^{-4 A'(y)}- \star_b
H_{w_1} \wedge E_{w_1} - \star_b H_{w_2} \wedge E_{w_2}\cr &
ds_{het}^2=\eta_{\m\n}dx^\m dx^\n+e^{-4A'(y)}
g_{ij}dy^idy^j+E_{w_1}E_{w_1}+E_{w_2}E_{w_2}
\end{split}
\end{equation}
supersymmtry will still be preserved. To this order in $\a'$ the
Bianchi identity becomes an equation of Laplace type, namely
\begin{equation}\label{app1}
d \star_b d e^{-4 A} - \star_b H_{w_k} \wedge H_{w_k} ={\a'\o 4}[\tr
(r\wedge r)-\tr ({\cal F}\wedge{\cal  F})].
\end{equation}

Note that we have obtained a linear differential for the dilaton
even though the Bianchi identity could, in principle, lead to a
highly non-linear differential equation. This fact depends crucially
on choosing the $\Omega_+$ connection to construct $\tr (R_+ \wedge
R_+)$. There is a preferred set of fields for which this connection
is required by space-time supersymmetry as shown by Bergshoeff and
de Roo \cite{Bergshoeff:1989de}. A different choice of connection is
always possible but it leads to a different choice of fields for
which in general the supersymmetry transformations will receive
corrections at $O(\a')$. We have found a differential equation of
Laplace type using the $\Omega_+$ connection and the solvability of
the equation is immediate if the integrated equation is satified.
Choosing the hermitian connection, on the other hand, will lead to a
highly non-linear differential equation of Monge-Ampere type as
shown in refs. \cite{Fu:2005sm}\cite{Fu:2006vj} .

In the following we will show that the $\a'$ corrected background
solves the equations of motion presented in section 3.2. First we
note that the equation of motion of ${\cal B}$ is satisfied since in
the background (\ref{corrected})
\begin{equation}\label{dddii}
\star_{10} {\cal H} = -e^{ 2 \ph} d\left(e^{-2 \ph} E^{w_1} \wedge
E^{w_2}\right) \wedge dx^{0123}.
\end{equation}
The Bianchi identity for ${\cal H}$ is solved by construction.

To solve the equations of motion for the metric we first establish
some properties of the Riemann tensor. First, the Ricci tensor of
the torsional metric is
\begin{equation}
R_{ij}=4 \nabla_i \partial _j A'+8 \partial_i A' \partial_j A' -
\half e^{4A'} H_{w_k a i } {H^{w_k a}}_j+g_{ij} \left[ 2 \nabla^2 A'
- 8 (\partial A')^2\right],
\end{equation}
where $(i,j)$ are indices on the base and $\nabla_i$ involves
connections on the base only. Note that this derivative is not
identical to $\nabla_i^{(6)}$, which is the covariant derivative
constructed with respect to the connections on the six-dimensional
torsional space. So for example
\begin{equation}
\nabla_i^{(6)} \partial_j \ph = \nabla_i \partial_j \ph - 8
\partial_i A'
\partial_j A' + 4 g_{ij} (\partial A')^2.
\end{equation}

Up to terms of $O(\a')$ the curvature two-form constructed from the
$\Omega_+$ connection ${R^A_+}_{B}$ satisfies
\begin{equation}
\star_b {R_+}^A{}_B=- {R_+}^A{}_B+O(\a')
\end{equation}
This condition can be derived using the integrability condition of
the supersymmetry constrain on the gravitino
\begin{equation}
\left[ {\nabla_-} _M , {\nabla_-}_ N\right] \e = {1\over 4}
{R_-}_{MNPQ} \Gamma^{PQ} \e=0,
\end{equation}
which implies
\begin{equation}
{R_-}_{MNP Q } J^{P  Q} =0.
\end{equation}
Moreover, one has
\begin{equation}
{R_-}_{PQMN}={R_+}_{MNPQ} - 2 \nabla_{[P } {\cal H}_{MNQ]} =
{R_+}_{MNPQ} +O(\a'),
\end{equation}
which implies
\begin{equation}
{R_+}_{P Q MN} J^{P  Q}+O(\a') =0.
\end{equation}
From here we obtain the following identity
\begin{equation}
{R_+}_{mPAB}{R_+}_{n}{}^{PAB}={1\o
4}{R_+}_{PQAB}{R_+}^{PQAB}g_{mn}+O(\a')
\end{equation}
where now $(m,n)$ are indices on the K3 base only, while if these
indices are along the fiber the result vanishes. Also,

\begin{equation}
\tr(R_+\wedge R_+)=-\half {R_+}_{PQAB}{R_+}^{PQAB}\star_b 1+O(\a').
\end{equation}

Using the above result for the curvature we can now verify the
equation of motion for the metric and the dilaton. The only
non-trivial component of the Einstein equation is the $(M,N)=(m,n)$
component with both indices along the base. All terms, except the
ones proportional to the base metric $g_{mn}$ cancel. The
coefficient of $g_{mn}$, on the other hand, turns out to be the
Hodge dual of the Bianchi identity (\ref{app1}), as can be verified
with a bit of patience. As a result the Einstein equation, Bianchi
identity and equation of motion for ${\cal B}$ are satisfied.
Explicit computation shows that also the dilaton equation of motion
is solved.

We end by describing torsional spaces with an N=2 supersymmetry in
which the twist of the fiber is `exchanged' by vacuum expectation
values of abelian gauge fields. This type of solutions were
suggested in refs. \cite{BTY:Heterotic}\cite{Sethi:2007bw}. In this
case the torus fiber is not twisted and the background fields are
\begin{equation}
\begin{split}
ds^2 & = \eta_{\m\n} dx^\m dx^\n + e^{-4 A'(y)} g_{ij} dy^i dy^j +
dw_1^2 + dw_2^2,\cr {\cal H} & =\star_bd e^{-4A'(y)}, \cr {\cal F} &
= {\cal F}_{i \bar j} dy^i dy^{\bar j},\cr \ph & = -2A'(y),
\end{split}
\end{equation}
where now an abelian gauge field is included as part of the
background and ${\cal F}$ is an anti-self dual form on K3. This
background solves the supersymmetry constraints preserving an N=2
supersymmtry. Moreover, it is not difficult to see that the Bianchi
identity reduces to the differential equation
\begin{equation}\label{nnii}
\begin{split}
-\nabla^2 e^{-4A(y)} \star_b 1= & {\a'\over 4} \left[ \tr (r \wedge
r) -\tr ({\cal F} \wedge {\cal F})\right] \cr &+  {3\a'\over 4}  d (
\nabla^2 e^{-4A} \star_b d e^{4A} ) \cr & +{\a'\over 4} d (e^{4A}
\star_b d \nabla^2 e^{-4A}).\cr
\end{split}
\end{equation}
The computation of $\tr (R_+ \wedge R_+)$ for these solutions is
greatly simplified since the fiber is not twisted. In this case the
second and third lines on the right hand side of eqn. (\ref{nnii})
are again corrections of order $O(\a'^2)$ or higher and can only be
consistently taken into account once the supersymmetry
transformations are corrected to $O(\a'^2)$. Therefore to $O(\a')$
the differential equation is again of Laplace type and solvability
is guaranteed. The form of the $O(\a'^2)$ corrections to the
supersymmetry transformations has been described in ref.
\cite{Bergshoeff:1989de}. The analysis to $O(\a'^2)$ of solutions
preserving an N=2 supersymmetry and the solvability of the Bianchi
identity for backgrounds preserving an N=1 supersymmetry is
currently under investigation \cite{begu}.

\section*{Acknowledgements}
It is our pleasure to thank Aaron Bergman, Greg Moore and Savdeep
Sethi for helpful discussions. K.~B. and G.~G. would like to thank
the KITP for hospitality during the final stages of this work.

\vskip 0.1 in \noindent This work is supported in part by NSF Grant
No. PHY-0505757 and by NSF Grant No. PHY-05-51164 and by the
University of Texas A\&M.

\appendix
\section{Conventions}

In this appendix, we summarize our conventions and quote some useful
formulas.

We use indices
$$M,N,
\dots \mu,\nu,\dots, i,j,\dots, w_1,w_2,\quad (A,B,\dots
\alpha,\beta,\dots, a,b,\dots , w_1 , w_2)$$ to denote the
coordinate (non-coordinate) bases of any six-dimensional space, of
four-dimensional Minkowski space-time, of the base and of the fiber,
respectively. For coordinates on the four-dimensional base of the
six-dimensional space, we use $y^i$ while we denote the fiber
coordinates by $w_i$, $i=1,2$.

We define the chirality operators
\begin{equation}
\Gamma^{(4)}=-i\G^0\G^1\G^2\G^3,\qquad
\G^{(4')}=-\G^4\G^5\G^6\G^7,\qquad\Gamma_\star=-i\G^8\G^9
\end{equation}
where $\G^{(4)}$, $\G^{(4')}$, and $\G_\star$ are the chirality
operators for external space, K3 base and the $T^2$ fibre, from
which we get
\begin{equation}
\Gamma^{(10)}=\Gamma^{(4)}\G^{(4')}\Gamma_\star=\G^0\cdots\G^9
\end{equation}
for the 10d space. In type the IIB theory, the 10d spinor $\e$
satisfies
\begin{equation}
\G^{(10)}\e=-\e
\end{equation}
We also choose the orientation
\begin{equation}
\ep^{4567w_1w_2}=-1.
\end{equation}
The Riemann tensor is defined by
\begin{equation}
R_{\m\n}{}^A{}_B=\p_\m \O^A{}_{B\n}-\p_\n
\O^A{}_{B\m}+\O^A{}_{C\m}\O^C{}_{B\n}-\O^A{}_{C\n}\O^C{}_{B\m}
\end{equation}
and

\begin{equation}
{\rm tr R\wedge R}=R^A{}_B\wedge R^A{}_B.
\end{equation}
We use the notation
\begin{equation}
H_{w_i}=\half H_{w_i ab}e^a\wedge e^b,\qquad (H_{w_i})_a=H_{w_i
ab}e^b,\qquad  H_{w_i ab}=H_{w_imn}e^m_ae^n_b
\end{equation}
and
\begin{equation}
|H|^2=\half H_{w_1ab}H_{w_1}{}^{ab}+\half H_{w_2ab}H_{w_2}{}^{ab}
\end{equation}
with $e^a$ the vielbein for unwarped K3 base.

To compute the ${\rm tr} R\wedge R$, it is convenient to use the
following results
\begin{equation}
\begin{split}
& A_{ij}A^i{}_k={1\o 4}A_{mn}A^{mn}g_{jk},  \qquad
S_{ij}S^i{}_k={1\o 4}S_{mn}S^{mn}g_{jk}\cr
&A_{ij}S^i{}_k=A_{ik}S^i{}_j,\qquad \qquad \qquad A_{ij}S^{ij}=0\cr
& A_{ij}=-\half\ep_{ij}{}^{kl}A_{kl},\qquad\qquad
S_{ij}=\half\ep_{ij}{}^{kl}S_{kl}
\end{split}
\end{equation}
where $A_{ij}$ are the components of any anti-self-dual two form on
the K3 base, and $S_{ij}$ are the components of any self-dual two
form on the K3 base.

\end{document}